\documentclass[prb,twocolumn,amsmath,amsfonts,floatfix,showpacs,a4paper]{revtex4}

\usepackage{graphicx}
\usepackage{epsfig}

\usepackage{color}

\begin{document}

\title{Anderson localisation in tight-binding models with flat bands}

\author{J. T. Chalker$^1$}
\author{T. S. Pickles$^1$}
\affiliation{$^1$Theoretical Physics, Oxford University, 1 Keble Road, Oxford, OX1 3NP, United Kingdom}
\author{Pragya Shukla$^{1,2}$}
\affiliation{$^2$Department of Physics, Indian Institute of Technology, Kharagpur, India}

\date{August 19, 2010}

\begin{abstract}
We consider the effect of weak disorder on eigenstates in a special class of tight-binding models.
Models in this class have short-range hopping on periodic lattices; their defining feature is that the clean systems have some energy bands that are dispersionless throughout the Brillouin zone.  We show that states derived from these flat bands are generically critical in the presence of weak disorder, being neither Anderson localised nor spatially extended. Further, we establish a mapping between this localisation problem and the one of resonances in random impedance networks, which previous work has suggested are also critical. 
Our conclusions are illustrated using numerical results for a two-dimensional lattice, known as the {square lattice with crossings} or the {planar pyrochlore lattice}.
\end{abstract}


\maketitle

\section{Introduction}
\label{introduction}

Hopping problems on certain frustrated lattices have the unusual feature that some of their energy bands are dispersionless. Systems of this kind have been studied in several different contexts, including both itinerant ferromagnetism within the Hubbard model \cite{mielke,tasaki} and frustrated antiferromagnetism within the Heisenberg model. \cite{chalker92,chalker-review} In the first of these examples, the divergent density of states for a flat band stabilises ferromagnetic order by ensuring that the Stoner criterion is satisfied. In the second example, the existence of a flat band in the hopping problem is closely linked to macroscopic ground state degeneracy in a classical antiferromagnet defined on the same lattice. 

The degeneracy of states in dispersionless bands is easily lifted by  perturbations. In this paper we examine the effect of one such perturbation: a weak random potential. For a flat band of the clean lattice, weak disorder sets the only energy scale and controls the localisation properties of eigenstates. Previous work in this area, \cite{goda-prl,nishino-jpsj} while identifying the weak-disorder limit as interesting, has focussed mainly on Anderson transitions occuring at finite disorder strength. We expect transitions occurring at finite disorder strength to be in the same universality class as those on unfrustrated lattices, and anticipate that special features of flat-band localisation occur only in the weak-disorder limit.

The weak-disorder flat-band localisation problem shares an obvious feature with the one of Anderson localisation in integer quantum Hall systems, in the sense that a Landau level can be viewed as a continuum analogue of a flat band. For a Landau level, a property of the projection operator onto the level -- its non-zero Chern number  -- ensures that disorder cannot localise all states.\cite{arovas}  Remarkably, we find for frustrated hopping problems that localisation is also controlled by a feature of the projection operator onto the flat band. In the cases we are concerned with, its matrix elements in real space decay as a power law of separation, and this power is equal to the spatial dimension. As a consequence, a random potential projected onto the flat band gives rise to a localisation problem similar to one in which the magnitude of hopping matrix elements between distant sites decreases as a power of their separation. A system of this kind is known to be critical when the power is the same as the spatial dimension, \cite{anderson,levitov} and the one-dimensional case of power-law banded random matrices has been studied extensively.\cite{power-law-banded} On these grounds, we expect eigenstates in frustrated tight binding models with weak disorder to be critical, and we use a numerical study of their multifractal properties to demonstrate this for a two-dimensional example. Separately, we show that the flat-band localisation problem can be mapped onto the problem of resonances occurring in random impedance networks, for which eigenfunctions have also been found to be critical in an earlier numerical study.\cite{luck}

While our exclusive focus here is on flat-band localisation, this problem has close links to a number of other topics of current interest. The most direct connection is to a geometrically frustrated antiferromagnet that has its spins polarised by a strong applied field: its single spin-flip excitations are described by a frustrated hopping Hamiltonian,\cite{schulenberg,zhitomirsky} and in the presence of weak disorder may exhibit the features we discuss in this paper (although the form of disorder we treat is site-diagonal, while randomness in the strength of exchange interactions would appear as off-diagonal disorder). More generally, the mathematics of flat bands is closely linked to the physics of Coulomb phases \cite{henley-review} in classical statistical mechanical systems -- notably dimer models \cite{huse} and frustrated magnets.\cite{isakov} There has recently been a general effort to understand transitions out of the Coulomb phase, induced by perturbations. \cite{alet,pickles,jaubert,andreanov,xu} In particular, flat-band localisation can be viewed as a linearised version of the transition induced by weak disorder in a frustrated magnet, between the Coulomb phase and spin glass order.\cite{andreanov} Finally, we note recent studies of Bose condensation in flat bands\cite{altman} and of the combined effects of disorder and interactions in the Falicov-Kimball model on a lattice with a flat band.\cite{souza}

In Section \ref{models} we define the models we consider, discuss projection onto the flat band and set out a mapping between the flat-band localisation problem and random impedance networks, while in Section \ref{simulations} we present results from simulations.

\section{Models and theoretical background}
\label{models}

Several parallel terminologies have been used to describe the tight-binding models we study here, which all involve a single non-zero hopping energy $t$ between nearby sites labelled $i$ at positions ${\bf r}_i$ on a periodic lattice in $d$ dimensions. In informal terms, for each model the sites can be grouped into clusters, labelled $\alpha, \beta \ldots$, in such a way that: $(i)$ each site is shared between two clusters; $(ii)$ every pair of sites belonging to a given cluster is linked by hopping; and $(iii)$ there is no hopping between sites from different clusters.
A second lattice appears naturally from this construction: it has its sites located at the centers of the clusters of the first lattice, and the sites of the first lattice lie at the mid-points of bonds of the second lattice. As described by Mielke \cite{mielke} using the language of graph theory, if the second lattice is denoted by the graph $G(V,E)$ with vertices $V$ and edges $E$, then the first lattice is the line graph $L(G)$ of $G$, which has as its vertex set $V_L$ the edge set $E$ of $G$, and has edges $E_L$ between two vertices in $V_L$ if and only if the corresponding edges in $G$ share a vertex. Our study is hence of tight binding models on line graphs. Alternatively, $G$ is known as the simplex lattice\cite{hassan} or parent lattice \cite{henley-review} and $L(G)$ is called its medial lattice. One distinction within this class is important to us: for reasons indicated below, we restrict ourselves to simplex lattices that are bipartite and expect different behaviour in the opposite case.  We note that some rather more general classes of tight binding models with flat bands have also been considered, \cite{nishino-flat} which we do not examine here.

\begin{figure}
\centering
\epsfig{file=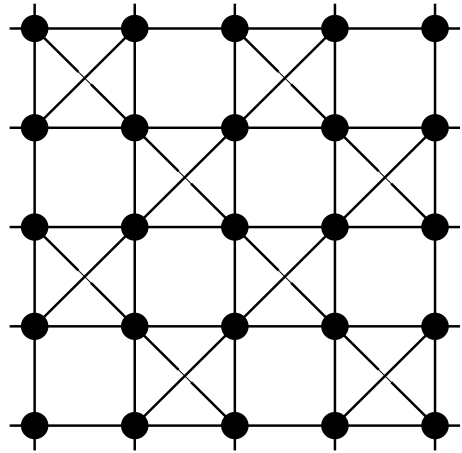,width=60mm}
\caption{\label{fig1} The square lattice with crossings, or planar pyrochlore lattice. Filled circles denote lattice sites and lines represent hopping matrix elements of magnitude $t$ in ${\cal H}^{(0)}$.}
\end{figure}

A two-dimensional example of the class of lattices which we are interested in is the square lattice with crossings, also called the planar pyrochlore lattice, shown in Fig.~\ref{fig1}. For this the simplex lattice is a square lattice. A second two-dimensional example is the kagome lattice, \cite{chalker92} which has the hexagonal lattice as its simplex lattice. A three-dimensional case is the pyrochlore lattice,\cite{chalker-review} with the diamond lattice as its simplex lattice; another is the octahedral lattice, \cite{henley-review} with the simple cubic lattice as its simplex lattice.

When specifying the tight-binding Hamiltonian ${\cal H}^{(0)}$ for the clean system it is convenient to include a constant term, chosen so that the flat bands are located at energy zero.
The Hamiltonian then has elements
\begin{equation}
{\cal H}^{(0)}_{ij} = \left\{
\begin{array}{ll}
t[1 + \delta_{ij}]& \,i,j \in {\rm same\, cluster}\\
0 & {\rm otherwise\,.}
\end{array}
\right.
\label{ham1}
\end{equation}
Its expectation value in a state with amplitudes $\varphi_i$ can conveniently be written in the form 
\begin{equation}
\langle \varphi | {\cal H}^{(0)} | \varphi \rangle = t \sum_\alpha \big| \sum_{i\in \alpha} \varphi_i \big|^2\,.  
\label{ham2}
\end{equation}
Taking $t>0$, it is clear from Eq.~(\ref{ham2}) that the eigenvalues of ${\cal H}^{(0)}$ are non-negative. It is also evident that states having 
\begin{equation}
\sum_{i\in \alpha} \varphi_i =0
\label{zero-energy}
\end{equation}
for all $\alpha$ and $\langle \varphi | \varphi \rangle \not= 0$ are eigenstates with energy zero. Such states can be constructed on the lattices we study by taking $\varphi_i = \pm 1$ with alternating signs on a closed loop of neighbouring sites that includes exactly two sites from every cluster visited by the loop, and setting $\varphi_i=0$ at sites not lying on the loop. (The construction with alternating signs requires the closed loop to have even length. Since the loop can be viewed as a walk on the simplex lattice, this property is guaranteed provided the simplex lattice is bipartite, but not otherwise.)  These loops can be chosen to be finite and short: for example, as squares on the planar pyrochlore lattice or as hexagons on the kagome lattice. Translated copies of such eigenstates form a band of zero-energy states, which is the flat band we are concerned with. 

In what follows, properties of the projection operator $\cal P$ onto the zero-energy states play an important role. When a band is separated by energy gaps from other bands, the real-space matrix elements  ${\cal P}_{ij}$ decay exponentially at long distances as a function of separation $|{\bf r}_i - {\bf r}_j|$ between sites.\cite{des-cloizeaux} However, for models in the class we have defined (here it is again crucial that the simplex lattice be bipartite), the zero-energy band touches a  dispersive band at one or more points in the Brillouin zone \cite{bergman} and ${\cal P}_{ij}$ decays more slowly with separation. For any specific lattice its form can be computed explicitly from the eigenvectors of ${\cal H}^{(0)}$, but to see the generic long-distance properties it is better to move to a continuum treatment. To this end, following discussions of the Coulomb phase in dimer models \cite{huse} and antiferromagnets, \cite{isakov} we first introduce real space unit vectors $\hat{e}_i$ aligned along each bond $i$ of the simplex lattice, and all directed from one sublattice (chosen as a matter of convention) towards the other. Then we  associate a $d$-component vector field ${\bf B}({\bf r}_i)$ with a set of amplitudes $\varphi_i$, by defining
\begin{equation}
{\bf B}({\bf r}_i) = \varphi_i \hat{e}_i\,.
\end{equation} 
The zero energy condition, Eq.~(\ref{zero-energy}), is also the condition that the field ${\bf B}({\bf r}_i)$ has zero lattice divergence at every site of the simplex lattice. To understand the long-distance behaviour of $\cal P$, we therefore consider the projection operator onto divergenceless fields that are functions of a continuous position variable $\bf r$. The reciprocal space form of $\cal P$ is block-diagonal in wavevector $\bf q$. On the lattice the blocks act on the space of sites within a unit cell, while in the continuum they are $d \times d$ matrices acting on the components of ${\bf B}$, with the form
\begin{equation}
\frac{q^2 \delta_{lm}- q_lq_m}{q^2}\,.
\end{equation}
In real space this yields
\begin{equation}
{\cal P}_{{\bf R}, {\bf R}+{\bf r}} = c_d \frac{d\,r_l r_m - r^2\delta_{lm}}{r^{d+2}}\,,
\end{equation}
where $c_d$ is a dimension-dependent constant. As advertised, the projection operator falls off as $r^{-d}$ in $d$-dimensions. 

Returning to the tight-binding model, we introduce disorder as a site-diagonal random potential with matrix elements
\begin{equation}
{\cal H}^{(1)}_{ij} =  \delta_{ij} v_i\,. 
\end{equation}
The $v_i$ are independent, identically distributed random variables having zero mean and unit variance.
Combining ingredients, the localisation problem we study has the Hamiltonian
\begin{equation}
{\cal H} = {\cal H}^{(0)} + {\cal H}^{(1)}\
\end{equation}
and we are interested in the weak disorder limit, obtained by taking $t \to \infty$.

Consider for $\cal H$ the density of states in energy $E$ when $t \gg 1$. At  $|E| \gg 1$ it is similar to that of ${\cal H}^{(0)}$, but for $|E|\lesssim {\cal O}(1)$ it is dominated by the contribution from the disorder-broadened flat band, which has energy width ${\cal O}(1)$. We wish to understand simplifications at large $t$  in the eigenvectors of $\cal H$ belonging to this band. Let
$|\varphi \rangle = |\varphi_{\parallel}\rangle + |\varphi_{\perp} \rangle$
be such an eigenvector, which we have separated into components that obey
${\cal P} |\varphi_{\parallel}\rangle = |\varphi_{\parallel}\rangle$
and $(1-{\cal P})|\varphi_{\perp} \rangle =|\varphi_{\perp} \rangle$
and lie respectively within and perpendicular to the space spanned by the flat bands of the clean system. Because ${\cal H}^{(0)}|\varphi_{\parallel}\rangle = 0$,
the eigenvalue equation $E|\varphi\rangle = {\cal H}|\varphi \rangle$ implies
\begin{equation}
(E-{\cal H}^{(1)} )|\varphi \rangle =  {\cal H}^{(0)}|\varphi_{\perp} \rangle\,.
\label{schroedinger}
\end{equation}
Since we expect $\left|(E-{\cal H}^{(1)} )|\varphi \rangle\right| \sim {\cal O}(1)$, we require $\left| |\varphi_{\perp} \rangle \right| \sim {\cal O}(t^{-1})$ to ensure that $\left| {\cal H}^{(0)} |\varphi_{\perp} \rangle \right| \sim {\cal O}(1)$ when $t$ is large. At leading order, Eq.~(\ref{schroedinger}) therefore simplifies to 
\begin{equation}
(E-{\cal H}^{(1)} )|\varphi_{\parallel} \rangle =  {\cal H}^{(0)}|\varphi_{\perp} \rangle\,.
\label{schroedinger2}
\end{equation}

To make a connection with random impedance networks, note that $\cal P$ and $(1-{\cal P})$ effect a lattice version of the Helmholtz decomposition: ${\cal P}$ projects onto lattice fields that are solenoidal, while $1-{\cal P}$ projects onto lattice fields that are irrotational. In consequence, ${\cal H}^{(0)}|\varphi_{\perp} \rangle$ can be expressed as the lattice gradient of a potential defined at the sites of the simplex lattice. Let $\alpha$ and $\beta$ be the simplex lattice sites linked by the simplex lattice bond $i$, with $\hat{e}_i$ directed from $\alpha$ to $\beta$. Then there exist simplex site potentials $\Phi_\alpha$ such that for every $i$
\begin{equation}
\langle i| {\cal H}^{(0)}|\varphi_{\perp} \rangle = \Phi_\alpha - \Phi_\beta\,.
\end{equation}
In addition
\begin{equation}
\langle i| (E - {\cal H}^{(1)}) |\varphi_\parallel \rangle = (E-v_i)\langle i | \varphi_\parallel \rangle
\end{equation}
and so with $w_{\alpha \beta}(E) = (E-v_i)^{-1}$ we have
\begin{equation}
\langle i | \varphi_\parallel \rangle = w_{\alpha \beta}(E) (\Phi_\alpha - \Phi_\beta)\,.
\end{equation}
Now Eq.~(\ref{zero-energy}) implies for every $\alpha$ that
\begin{equation}
\sum_{\beta} w_{\alpha \beta}(E) (\Phi_\alpha - \Phi_\beta) = 0\,
\label{potential}
\end{equation}
where the sum on $\beta$ is over simplex lattice sites that are nearest neighbours to $\alpha$. 

Eq.~(\ref{potential}) is also the equation that describes resonances in an impedance network.\cite{luck} It has non-trivial solutions for $\Phi_\alpha$ only if $E$ is an eigenvalue of $\cal H$. In the impedance network interpretation, it is Kirchoff's law for current conservation at node $\alpha$ of the network, $\Phi_\alpha$ is the voltage at that node, and $w_{\alpha \beta}(E) $ is the complex conductance, or inverse impedance, between nodes $\alpha$ and $\beta$. This impedance is frequency-dependent and the equation has non-trivial solutions at the resonant frequencies, which are real if the network is loss-free. Such resonances have been studied for a binary distribution of impedances on the square lattice by Jonckheere and Luck.\cite{luck} In particular, these authors find from numerical calculations that the electric fields associated with resonant states, which correspond in our notation to the combination  $(E-v_i)\langle i | \varphi_\parallel \rangle$, have a multifractal distribution. Their conclusion is entirely consistent with our view of flat-band localisation, as being automatically critical because $\cal P$ is long range. Moreover, this connection to flat-band localisation provides an explanation of why resonant states should generically be critical for an impedance network in any number of dimensions.

\section{Numerical results}
\label{simulations}

To illustrate these ideas and study the approach to the large $t$ limit, we have calculated eigenfunctions and eigenvectors for $\cal H$ on the planar pyrochlore lattice of Fig.~\ref{fig1}. We take the site potentials $v_i$ from a Gaussian distribution and examine values of $t$ up to $3 \times 10^9$. We diagonalise $H$ for square systems with periodic boundary conditions and side length $L$ in the range $34\leq L \leq 80$, in units of the nearest neighbour spacing.  

\begin{figure}
\centering
\epsfig{file=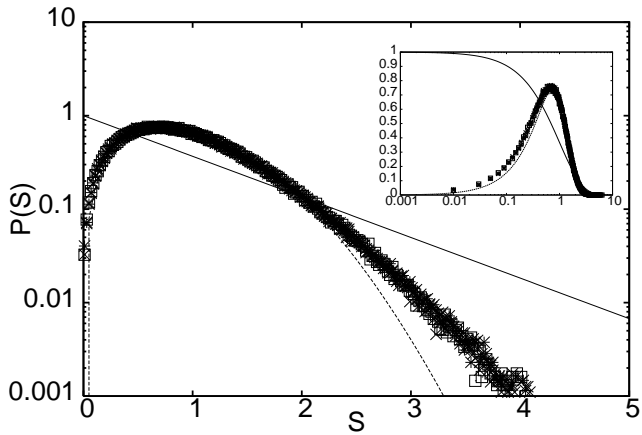,angle=270,width=85mm}
\caption{\label{fig2} The probability distribution of nearest-neighbour level spacings $P(s)$, obtained for system sizes $L=34$ ($\times$), 44 ($*$) and 54 ($\Box$). Inset: detailed view of behaviour at small $s$. The Poisson distribution is represented by the full line, and the Wigner-Dyson distribution by the dashed line.}
\end{figure}

The probability distribution $P(s)$ of spacings $s$ between adjacent eigenvalues, measured in units of the mean spacing, is a standard diagnostic for localisation transitions.\cite{altshuler-shklovskii,shklovskii} Metallic and localised phases are identified by Wigner-Dyson and Poisson distributions\cite{mehta} respectively, while the critical point is characterised by a distinct, universal distribution.\cite{shklovskii} We present in Fig.~\ref{fig2} our results for $P(s)$. These were obtained using states from the central half of the flat band, and applying standard band-unfolding methods to compensate for the energy dependence of the average density of states. The data shown were calculated using $t=3 \times10^9$ but results are essentially the same for any $t\geq 10^3$. Within the statistical precision of the data, the behaviour of $P(s)$ is independent of system size. It is apparent from this figure that the form of $P(s)$ is intermediate between those for Wigner-Dyson and Poisson distributions, as expected in a system at a critical point. 

\begin{figure}[t]
\centering
\epsfig{file=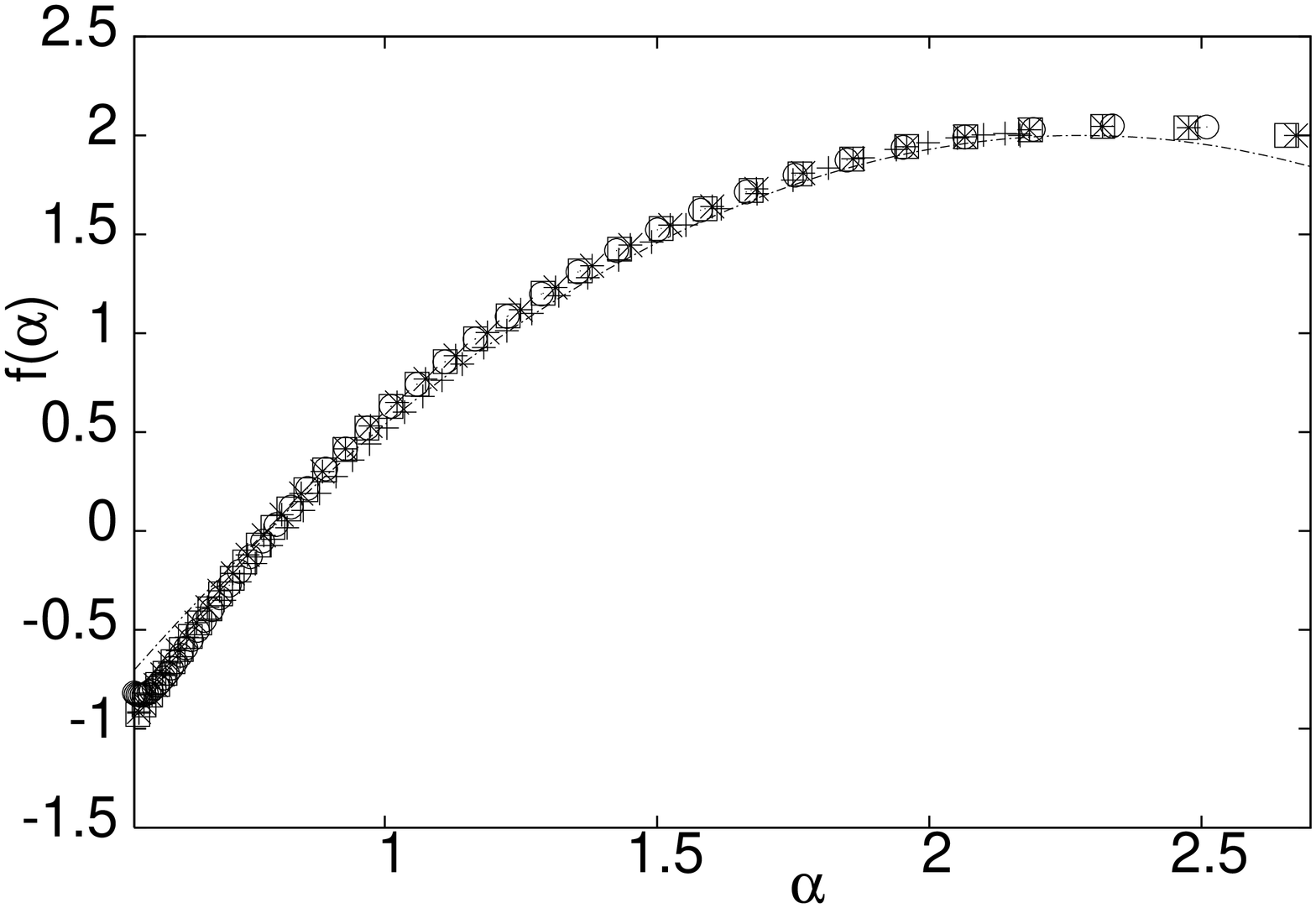,angle=270,width=85mm}
\epsfig{file=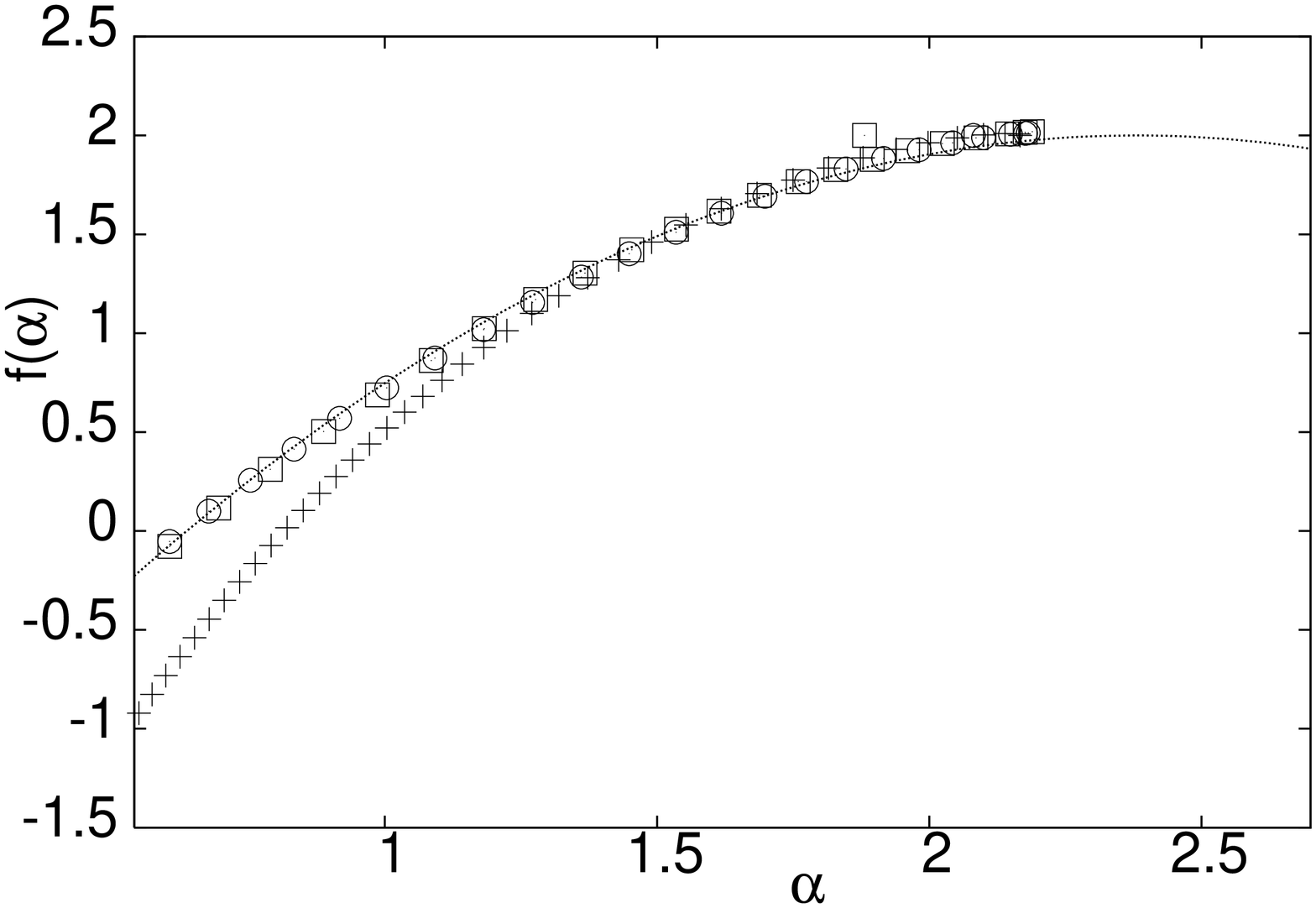,angle=270,width=85mm}
\caption{\label{fig3} The singularity spectrum $f(\alpha)$ determined for eigenstates from the disorder-broadened flat band. {\it Upper panel:} Size-dependence of results computed for $t=3 \times10^3$; data are for  $L=34$ ($\circ$), 54 ($\Box$), and 80 ($\times$); $+$ indicates an extrapolation to $L=\infty$ and the dashed line is a parabolic fit to this extrapolation. {\it Lower panel:} $t$-dependence of extrapolations for $t = 3 \times10^3$ ($\times$), $3 \times 10^5$ ($\Box$), and $3 \times10^8$ ($\circ$). The dashed line is a parabolic fit to the extrapolation for the largest value of $t$.}
\end{figure}

To probe the nature of eigenfunctions we have examined the scaling with system size of averaged moments of the probability density. More specifically, eigenfunction fluctuations can be characterised by a set of generalised fractal dimensions \cite{janssen} if disorder-averaged moments have the power-law dependence
\begin{equation}
\sum_i [{ |\varphi_i|^{2q}}]_{\rm av}  = A_q L^{-\tau(q)}\,
\end{equation}
on system size, where $\tau(q)$ is related to the fractal dimension by $\tau(q) = (q-1)D_q$ and $A_q$ is a constant. For uniformly extended states $D_q=d$ and for localised states $D_q=0$, but for critical states a non-trivial dependence of $D_q$ on $q$ is expected. As is standard, \cite{janssen} we analyse multifractality via the spectrum of singularity strengths $f(\alpha)$, which is related to $\tau(q)$ by a Legendre transformation: $f(\alpha) = q\alpha - \tau(q)$ with $q = \partial_\alpha f(\alpha)$. It has the interpretation that $L^{f(\alpha)}$ is the measure of the set of points $i$ at which the probability density in an eigenstate scales with system size as $|\varphi_i|^2 \sim L^{-\alpha}$. We determine $f(\alpha)$ using the procedure described in Ref.~\onlinecite{evers}. The results are shown in Fig.~\ref{fig3}. The upper panel of this figure displays the size-dependence of our estimates for $f(\alpha)$ at finite $L$, showing convergence to the large-$L$ limit. The lower panel demonstrates that the form of $f(\alpha)$ that results at large $L$ is $t$-independent provided $t$ is large.\cite{footnote} These final, large-$L$, large-$t$ data are well-fitted by the parabola
\begin{equation}
f(\alpha) = d - \frac{(\alpha - \alpha_0)^2}{4(\alpha_0 - d)}
\end{equation} 
with $d=2$ and $\alpha_0 = 2.37$. Correspondingly, we find $D_2 = 1.27$. 
Fractal properties in generalized random matrix ensembles depend
on details of the model\cite{shukla} (for example, on disorder strength in power law random banded matrices\cite{power-law-banded}).
Similarly, we expect fractal properties in weak-disorder flat-band localisation to vary with the choice of distribution for $v_i$, but we have not explored that aspect.

In summary, these numerical results support our expectation that weak disorder in a tight binding model with flat bands should give rise to eigenstates that are critical, in the sense of the Anderson localisation transition.
 
\begin{acknowledgments}
This work was supported in part by the Royal Society and by EPSRC Grant No. EP/D050952/1.  
\end{acknowledgments}


\begin{thebibliography}{99}
\bibitem{footnote} Strictly, we do not expect the limits $t\to\infty$ and $L\to \infty$ to commute: at fixed $t$ there should be a crossover
at sufficiently large $L$ to a conventional localisation problem. Our results indicate that the scale at which this happens 
must be very large when $t$ is large. We therefore believe that our determination of $f(\alpha)$ is unaffected by the crossover.
\end{thebibliography}
\end{document}